# Pressure-induced Superconductivity at 32 K in $MoB_2$

Cuiying Pei[1#], Jianfeng Zhang[2#], Qi Wang[1,2,3#], Yi Zhao[1], Lingling Gao[1], Chunsheng Gong[2], Shangjie Tian[2], Ruitao Luo[2], Zhong-Yi Lu[2], Hechang Lei[2*], Kai Liu[2*], and Yanpeng Qi[1*]

[1]School of Physical Science and Technology, ShanghaiTech University, Shanghai 201210, China

[2]Department of Physics and Beijing Key Laboratory of Opto-electronic Functional Materials & Micro-nano Devices, Renmin University of China, Beijing 100872, China

[3]ShanghaiTech Laboratory for Topological Physics, ShanghaiTech University, Shanghai 201210, China

**Abstract: Since the discovery of superconductivity in $MgB_2$ ($T_c \sim 39$ K), the search for superconductivity in related materials with similar structures or ingredients has never stopped. Although about 100 binary borides have been explored, only few of them show superconductivity with relatively low $T_c$. In this work, we report the discovery of superconductivity up to 32 K in $MoB_2$ under pressure which is the highest $T_c$ in transition-metal borides. Although the $T_c$ can be well explained by theoretical calculations in the framework of electron-phonon coupling, the *d*-electrons and phonon modes of transition metal Mo atoms play utterly important roles in the emergence of superconductivity in $MoB_2$, distinctly different from the case of well-known $MgB_2$. Our study sheds light on the exploration of high-$T_c$ superconductors in transition metal borides.**

Superconductors with high transition temperature (high-$T_c$) are long-sought targets in the condensed matter physics and materials communities. Materials with light elements[1-5] are especially favorable as they can provide high Debye frequency, which is proportional to the superconducting $T_c$ according to the Bardeen-Cooper-Schrieffer (BCS) theory[6]. In various light-element materials, metal borides have attracted much attention due to their unique crystal structure and rich physical phenomena. The discovery of superconductivity in $MgB_2$ at 39 K in 2001 has reinforced the scientific importance of metal diborides ($MB_2$)[1]. Nevertheless, after the exploration of superconductivity in the rich family of metal diborides for nearly two decades, only few of $MB_2$ show superconducting behavior. Furthermore, the $T_c$s of these superconducting $MB_2$ compounds are much lower than that of $MgB_2$ (Table 1 and Table S1).

Molybdenum diboride ($MoB_2$) is unique among the $MB_2$ family since it is the only material that has two structural forms, $\alpha$-$MoB_2$ phase ($AlB_2$-type, space group $P6/mmm$)[7-9] and $\beta$-$MoB_2$ phase ($CaSi_2$-type, space group $R\bar{3}m$)[10-12]. Although both phases of $MoB_2$ have similar triangular Mo layers, their arrangements of B atoms are quite different. The B atoms in $\alpha$-$MoB_2$ constitute the AA-staking two-dimensional (2D) graphitic boron layers. Instead, there are two different kinds of B layers in $\beta$-$MoB_2$ (Fig. 1a): one forms a nearly planar quasi-2D honeycomb lattice similar to those in $\alpha$-$MoB_2$ and $MgB_2$; the other builds a buckled honeycomb network. $MoB_2$ may thus provide a good material platform to make a comparative study with $MgB_2$.

Pressure, as a conventional thermodynamic parameter, is a clean and powerful tool to tune the electronic properties of materials. It is possible to trigger structural and electronic transitions, subsequently, inducing novel quantum phenomena[13]. Here, we report on the structural and transport properties of $\beta$-$MoB_2$ at various pressures. We find that $MoB_2$ exhibits superconductivity under high pressure, reaching a $T_c$ of 32 K around 100 GPa, which is the second highest $T_c$ among all known boride superconductors. Synchrotron X-ray diffraction (XRD) measurements indicate that $\beta$-$MoB_2$ transforms to $\alpha$-$MoB_2$ around 65 GPa. Theoretical calculations demonstrate the important roles of Mo $d$-electrons and Mo-derived phonon modes playing in the high-$T_c$ superconductivity of $\alpha$-$MoB_2$ under pressure, which is distinctly different from the mechanism in $MgB_2$.

High-quality single crystals of $MoB_2$ were grown by Al flux. The crystalline phases used in high-pressure experiments were identified by powder x-ray diffraction (XRD) and inductive coupled plasma (ICP) optical emission spectroscopy (OES). *In situ* high pressure XRD and resistivity measurements were performed in a diamond anvil cell (DAC). We employed the swarm-intelligence-based CALYPSO structure prediction method[14] to find the energetically stable structures of $MoB_2$ under high pressure. The electronic structure, phonon spectrum, and electron-phonon coupling (EPC) of $AlB_2$-type $\alpha$-$MoB_2$ at 90 GPa were studied based on the density functional theory (DFT)[15,16] and density functional perturbation theory (DFPT)[17,18] calculations. The details of experiment are shown in Supplemental Information.

We performed *in situ* XRD measurements on the structural evolution of $MoB_2$ under various pressures. As shown in Fig. 1b, all the diffraction peaks in the low-pressure range can be indexed well to a rhombohedral primitive cell of $\beta$-$MoB_2$ (space group $R\bar{3}m$), and both $a$-axis and $c$-axis lattice constants decrease with increasing pressure (Supplemental Information Figs. S1-S2 and Table S2). The structure of $\beta$-$MoB_2$ is robust until 65 GPa. Beyond ~65 GPa, additional diffraction peaks emerge, indicating an occurrence of structural phase transition. Meanwhile, we carried out a global minimization of the free energy of $MoB_2$ under high pressure by combining *ab initio* total energy calculations and the Calypso technique on structure predictions[14]. At 90 GPa, we theoretically predict two structural forms (space groups $P6/mmm$ and $I4_1/amd$) of $MoB_2$ with lower enthalpies than the $R\bar{3}m$ structure (Supplemental Information Fig. S3a). No imaginary frequency is found in the phonon dispersions of these two predicted structures, suggesting their dynamical stability (Supplemental Information Figs. S3c and S3d). It is found that the XRD pattern at 90 GPa can be well refined by using the hexagonal $\alpha$-$MoB_2$ structure (space group $P6/mmm$) (Supplemental Information Fig. S3b and Fig. S4). These experimental and theoretical results suggest that under high pressure there is a structural phase transition from $\beta$-$MoB_2$ to $\alpha$-$MoB_2$ with a critical pressure $P_c$ ~ 70 GPa.

Since $MoB_2$ under high pressure possesses the same crystal structure as $MgB_2$, a question arises naturally: is it possible to achieve superconductivity in $MoB_2$ under high pressure? Hence, we measured electrical resistivity $\rho(T)$ of $\beta$-$MoB_2$ single crystal at various pressures. Figure 2a shows the typical $\rho(T)$ curves for pressure up to 109.7 GPa. The $\rho(T)$ curves display a metallic-like behavior in the whole pressure range. When the pressure increases to 21.7 GPa, a small drop of $\rho$ is observed at the lowest measuring temperature ($T_{min}$ = 1.8 K), as shown in Fig. 2b. With further increasing pressure, zero resistivity is achieved at low temperature for $P > 38.5$ GPa, indicating the emergence of superconductivity. The superconducting $T_c$ increases dramatically with pressure. Beyond the critical pressure ($P_c = 70$ GPa) where the structural phase transition happens, the growth of $T_c$ slows down and the maximum $T_c$ of 32.4 K is attained at $P = 109.7$ GPa, which is the highest pressure we can exert on the sample. At this pressure, the superconducting transition width $\Delta T_c$ [$= T(90\%\ \rho_n) - T(10\%\ \rho_n)$] is about 4.2 K (Fig. 2b), which is rather sharp when compared with the large value of $T_c$. Figure 2c demonstrates that at 109.7 GPa the resistivity drop is continuously suppressed with increasing magnetic field and it shifts to about 5 K at 9 T. Such behavior further confirms that the sharp decrease of $\rho(T)$ should originate from a superconducting transition. The derived upper critical field $\mu_0H_{c2}(T)$ as a function of temperature $T$ can be fitted well using the empirical Ginzburg-Landau formula (Fig. 2d) $\mu_0H_{c2}(T) = \mu_0H_{c2}(0)(1-t^2)/(1+t^2)$, where $t = T/T_c$ is the reduced temperature with zero-field superconducting $T_c$. The fitted zero-temperature upper critical field $\mu_0H_{c2}(0)$ of $MoB_2$ from the 90% $\rho_n$ criterion can reach 9.4(1) T with $T_c$ = 31.7(2) K, which is consistent with the measured value of 32.4 K.

The measurements on different samples of $MoB_2$ for five independent runs provide the consistent and reproducible results (Supplemental Information Fig. S5), confirming this intrinsic superconductivity under pressure. The superconducting phase diagram of $MoB_2$ as a function of pressure is summarized in Fig. 3. It can be seen that the superconducting state emerges around 20 GPa, and then the $T_c$ increases further with applied pressure. The $T_c$ raises dramatically at a rate of 0.7 K/GPa in the range of 40 to 70 GPa, and beyond the structure-transition pressure ($P_c \sim 70$ GPa) the growth of $T_c$ slows down (0.1 K/GPa). The $T_c$ of $MoB_2$ rises to as high as ~ 32 K at the pressure of 109.7 GPa and still does not exhibit the trend of saturation.

To explore the origin of the relatively high-$T_c$ superconductivity in $MoB_2$ under pressure, we performed the density functional theory (DFT)[15,16] and density functional perturbation theory (DFPT)[17,18] calculations on $MoB_2$ in $\alpha$-phase at 90 GPa, for which the experimental $T_c$ can reach ~ 31.5 K at 91.4 GPa. The calculated band structure of $\alpha$-$MoB_2$ at 90 GPa along the high-symmetry paths of the Brillouin zone is shown in Fig. 4a. There are several bands crossing the Fermi level ($E_F$), indicating its metallic character. Based on the analysis of total and partial density of states (Fig. 4b), we can learn that Mo $d$-orbitals (especially the $d_{z2}$ orbital) have larger contributions than B $p$-orbitals around the $E_F$. The phonon dispersion and the phonon density of states $F(\omega)$ for $\alpha$-$MoB_2$ at 90 GPa are displayed in Figs. 4c and 4d, respectively. Due to the large mismatch between the atomic masses of Mo and B, there is a gap between the low-frequency acoustic branch contributed by Mo atoms and the high-frequency optical branch contributed by B atoms. Obviously, the acoustic modes along the A-L path and around the L and H points make great contributions to the electron-phonon coupling (EPC, represented by the red dots in Fig. 4c), which is also manifested in the Eliashberg spectral function $\alpha^2F(\omega)$ (red line in Fig. 4d). Figure 4e demonstrates a representative phonon mode of $\alpha$-$MoB_2$ at the H point at 90 GPa. For clear exhibition, a $\sqrt{3} \times \sqrt{3} \times 2$ supercell is employed and the arrows on B atoms are magnified by two times. It can be seen that this phonon mode consists of the relative vibrations of Mo atoms perpendicular to the B-B honeycomb plane and the related in-plane breath-like vibrations of B atoms. The calculated total EPC constant $\lambda$ is 1.60. Based on the McMillan-Allen-Dynes formula[19,20], we can then calculate the superconducting $T_c$ with an effective screened Coulomb repulsion constant $\mu^*$. By adopting several values of $\mu^*$ in a commonly-used empirical range of 0.08 to 0.15[19,20], we obtained the $T_c$ between 27.2 K and 33.3 K (see FIG. S7 in Supplemental Material), showing good order of magnitude with the measured one (~32 K). These theoretical calculations confirm that the observed superconductivity in $MoB_2$ at high pressure should belong to the conventional BCS type.

Although $\alpha$-$MoB_2$ at high pressure shares the same crystal structure as $MgB_2$ and owns a comparable superconducting $T_c$ with that of $MgB_2$ at ambient pressure[1], their electronic structures and superconducting ingredients are distinctly different. Firstly, for $MgB_2$, the $p$ orbitals of B atoms in the B-B honeycomb lattice contribute most around the Fermi level[21,22], while for $\alpha$-$MoB_2$ under high pressure the states near the $E_F$ are dominated by the $d$-electrons of Mo atoms (Fig. 4b). As a result, $MoB_2$ has very large band dispersion along the $k_z$ direction (such as the Γ-A path) (Fig. 4a) and its Fermi surface shows three-dimensional characteristic, i.e. without such two-dimensional Fermi surface as in $MgB_2$[22,23] (Supplemental Information Fig. S6). Secondly, the main contribution to the EPC of $\alpha$-$MoB_2$ under high pressure derives from the low-frequency phonon branch (Figs. 4c and 4d); instead the high-frequency branch plays a major role in $MgB_2$[24-27]. To be specific, for high-pressure $\alpha$-$MoB_2$ the out-of-plane phonon mode of Mo atoms couples strongly with Mo $d$-electrons near the $E_F$ (Figs. 4b and 4e). In comparison, it is the in-plane B-B stretching mode in $MgB_2$ interacts intensively with the $\sigma$-bond in the boron honeycomb lattice around the $E_F$ (Fig. 4f)[21]. Last but not least, previous theoretical studies suggest that the anisotropy in the EPC and the anharmonicity in the stretching phonon mode are very crucial for the high $T_c$ of $MgB_2$[26-28]. Here for $\alpha$-$MoB_2$ under high pressure, our calculations seem to coincide with the observed $T_c$ without invoking above two factors. These results reveal that the superconducting mechanism in high-pressure $\alpha$-$MoB_2$ is distinct from that in $MgB_2$, suggesting the probability of exploring new phonon-mediated high-$T_c$ superconductors in transition metal borides.

In addition to the $\alpha$-phase of $MoB_2$, we have also calculated the superconducting $T_c$ of $\beta$-$MoB_2$ by using the same method. Nevertheless, the calculated $T_c$ of $\beta$-$MoB_2$ is always lower than 5 K and shows a decreasing tendency with pressure (see FIG. S9 in Supplemental Material), which are inconsistent with our experimental observations (FIG. 3). In consideration of the helical nodal lines in the phonon spectrum previously reported for $\beta$-$MoB_2$[12], whether or not the special topological phonon dispersions will have some effect on its superconductivity remains unknown. The origin of superconductivity in $\beta$-$MoB_2$ at low pressures waits for further experimental and theoretical investigations in the future.

In summary, we found superconductivity up to 32 K in $MoB_2$ under pressure. Although compressed $\alpha$-$MoB_2$ and $MgB_2$ both possess the $AlB_2$-type structures and comparable $T_c$, the features of their electron-phonon couplings that mediate the superconductivity are quite different. This resembles the situation in recently reported lanthanum hydride superconductor $LaH_{10}$ with $T_c \approx 250 - 260$ K[4,29-31], where the strong EPC are contributed by the coupling between the hybridized La $f$ – H $s$ electrons and the high-frequency phonon modes of H atoms[32]. Thus, not only do the electrons in light-element atom networks need to be considered, but also the relatively localized electrons of metal atoms may be worthy of attention in the search for high-$T_c$ superconductors.

[#] These authors contributed to this work equally.

[*] Correspondence should be addressed to Y.Q. (qiyp@shanghaitech.edu.cn) or K.L. (kliu@ruc.edu.cn) or H.C.L. (hlei@ruc.edu.cn)

**Acknowledgment**

The authors thank the support from Analytical Instrumentation Center (# SPST-AIC10112914), SPST, ShanghaiTech University. The authors thank the staffs from BL15U1 at Shanghai Synchrotron Radiation Facility for assistance during data collection. This work was supported the National Key R&D Program of China (Grant No. 2018YFA0704300, 2018YFE0202600, 2017YFA0302903, and 2016YFA0300504), the National Natural Science Foundation of China (Grant No. U1932217, 11774424, 11774423, 11822412, and 11974246), the Natural Science Foundation of Shanghai (Grant No. 19ZR1477300), the Beijing Natural Science Foundation (Grant No. Z200005), the CAS Interdisciplinary Innovation Team, the Fundamental Research Funds for the Central Universities, and the Research Funds of Renmin University of China (Grant No. 18XNLG14, 19XNLG13, and 19XNLG17). Computational resources were provided by the Physical Laboratory of High Performance Computing at Renmin University of China.

**References**

[1] J. Nagamatsu, N. Nakagawa, T. Muranaka, Y. Zenitani, and J. Akimitsu, Nature **410**, 63 (2001).
[2] J. R. Tobin, *Superconductivity research developments* (Nova Science Pub Inc, 2008).
[3] A. P. Drozdov, M. I. Eremets, I. A. Troyan, V. Ksenofontov, and S. I. Shylin, Nature **525**, 73 (2015).
[4] A. P. Drozdov *et al.*, Nature **569**, 528 (2019).
[5] E. Snider, N. Dasenbrock-Gammon, R. McBride, M. Debessai, H. Vindana, K. Vencatasamy, K. V. Lawler, A. Salamat, and R. P. Dias, Nature **586**, 373 (2020).
[6] J. Bardeen, L. N. Cooper, and J. R. Schrieffer, Phys. Rev. **108**, 1175 (1957).
[7] K. Kudaka, K. Iizumi, T. Sasaki, and S. Okada, J. Alloys Compd. **315**, 104 (2001).
[8] R. Steinitz, I. Binder, and D. Moskowitz, JOM **4**, 983 (1952).
[9] L.-P. Ding, P. Shao, F.-H. Zhang, C. Lu, L. Ding, S. Y. Ning, and X. F. Huang, Inorg. Chem. **55**, 7033 (2016).
[10] S. Okada, T. Atoda, I. Higashi, and Y. Takahashi, J. Mater. Sci. **22**, 2993 (1987).
[11] M. Frotscher, W. Klein, J. Bauer, C.-M. Fang, J.-F. Halet, A. Senyshyn, C. Baehtz, and B. Albert, Z. Anorg. Allg. Chem. **633**, 2626 (2007).
[12] T. T. Zhang *et al.*, Phys. Rev. Lett. **123**, 245302 (2019).
[13] Y. Qi *et al.*, Nat. Commun. **7**, 11038 (2016).
[14] Y. Wang, J. Lv, L. Zhu, and Y. Ma, Comput. Phys. Commun. **183**, 2063 (2012).
[15] P. Hohenberg and W. Kohn, Phys. Rev. **136**, B864 (1964).
[16] W. Kohn and L. J. Sham, Phys. Rev. **140**, A1133 (1965).
[17] F. Giustino, Rev. Mod. Phys. **89**, 015003 (2017).
[18] S. Baroni, S. de Gironcoli, A. Dal Corso, and P. Giannozzi, Rev. Mod. Phys. **73**, 515 (2001).
[19] K. H. Lee, K. J. Chang, and M. L. Cohen, Phys Rev B Condens Matter **52**, 1425 (1995).
[20] C. F. Richardson and N. W. Ashcroft, Phys. Rev. Lett. **78**, 118 (1997).
[21] J. M. An and W. E. Pickett, Phys. Rev. Lett. **86**, 4366 (2001).
[22] J. Kortus, I. I. Mazin, K. D. Belashchenko, V. P. Antropov, and L. L. Boyer, Phys. Rev. Lett. **86**, 4656 (2001).
[23] H. J. Choi, D. Roundy, H. Sun, M. L. Cohen, and S. G. Louie, Nature **418**, 758 (2002).
[24] K. P. Bohnen, R. Heid, and B. Renker, Phys. Rev. Lett. **86**, 5771 (2001).
[25] Y. Kong, O. V. Dolgov, O. Jepsen, and O. K. Andersen, Physical Review B **64**, 020501 (2001).
[26] A. Y. Liu, I. I. Mazin, and J. Kortus, Phys. Rev. Lett. **87**, 087005 (2001).
[27] T. Yildirim *et al.*, Phys. Rev. Lett. **87**, 037001 (2001).
[28] H. J. Choi, D. Roundy, H. Sun, M. L. Cohen, and S. G. Louie, Phys. Rev. B **66**, 020513 (2002).
[29] M. Somayazulu, M. Ahart, A. K. Mishra, Z. M. Geballe, M. Baldini, Y. Meng, V. V. Struzhkin, and R. J. Hemley, Phys. Rev. Lett. **122**, 027001 (2019).
[30] F. Peng, Y. Sun, C. J. Pickard, R. J. Needs, Q. Wu, and Y. Ma, Phys. Rev. Lett. **119**, 107001 (2017).
[31] H. Liu, I. I. Naumov, R. Hoffmann, N. W. Ashcroft, and R. J. Hemley, Proc. Natl Acad. Sci. USA **114**, 6990 (2017).
[32] L. Liu, C. Wang, S. Yi, K. W. Kim, J. Kim, and J.-H. Cho, Phys. Rev. B **99**, 140501 (2019).
[33] C. Buzea and T. Yamashita, Supercond. Sci. Technol. **14**, R115 (2001).
[34] B. T. Matthias, E. Corenzwit, J. M. Vandenberg, H. Barz, and -. 74, Proc. Natl. Acad. Sci. USA **74**, 1334 (1977).
[35] H. C. Ku and R. N. Shelton, Mat. Res. Bull. **15**, 1441 (1980).
[36] R. N. Shelton and H. C. Ku, Mat. Res. Bull. **15**, 1445.

[37] P. Lejay, B. Chevalier, J. Etourneau, P. Hagenmuller, and P. Peshev, Synth. Met. **4**, 139 (1981).
[38] B. T. Matthias, T. H. Geballe, K. Andres, E. Corenzwit, G. W. Hull, and J. P. Maita, Science **159**, 530 (1968).
[39] H. E. E., D. H., and K. J. M., J. Less-Common Met. **27**, 281 (1972).
[40] R. N. Shelton, J. Less-Common Met. **62**, 191 (1978).
[41] A. Kawano, Y. Mizuta, H. Takagiwa, T. Muranaka, and J. Akimitsu, J. Phys. Soc. Japan **72**, 1724 (2003).
[42] H. C. Ku, G. P. Meisner, F. Acker, and D. C. Johnston, Solid State Commun. **35**, 91 (1980).
[43] T. Sakai, G.-Y. Adachi, and J. Shiokawa, J. Less-Common Met. **84**, 107 (1982).

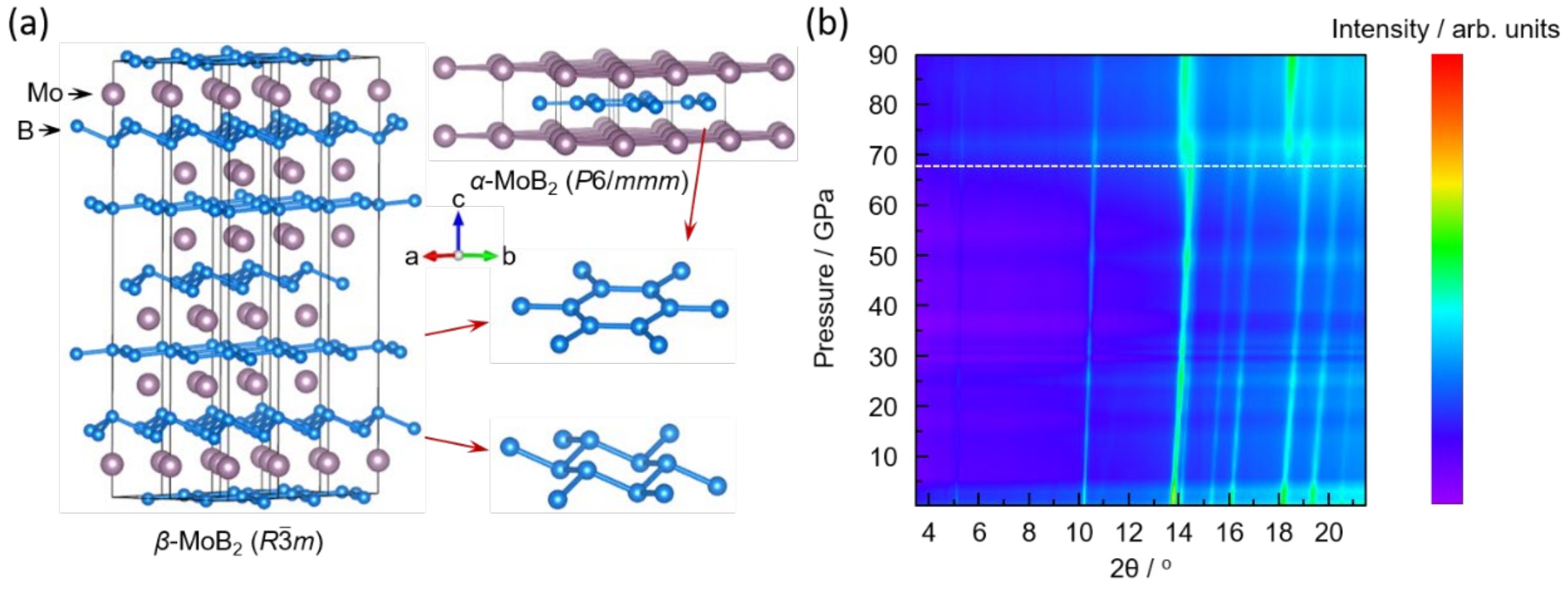

FIG. 1 (color online). Crystal structure of $MoB_2$ and its evolution with pressure. (a) Crystal structures of $\alpha$-$MoB_2$ (space group $P6/mmm$) and $\beta$-$MoB_2$ (space group $R\bar{3}m$). (b) Contour color plot of the pressure-dependent XRD patterns up to 90 GPa.

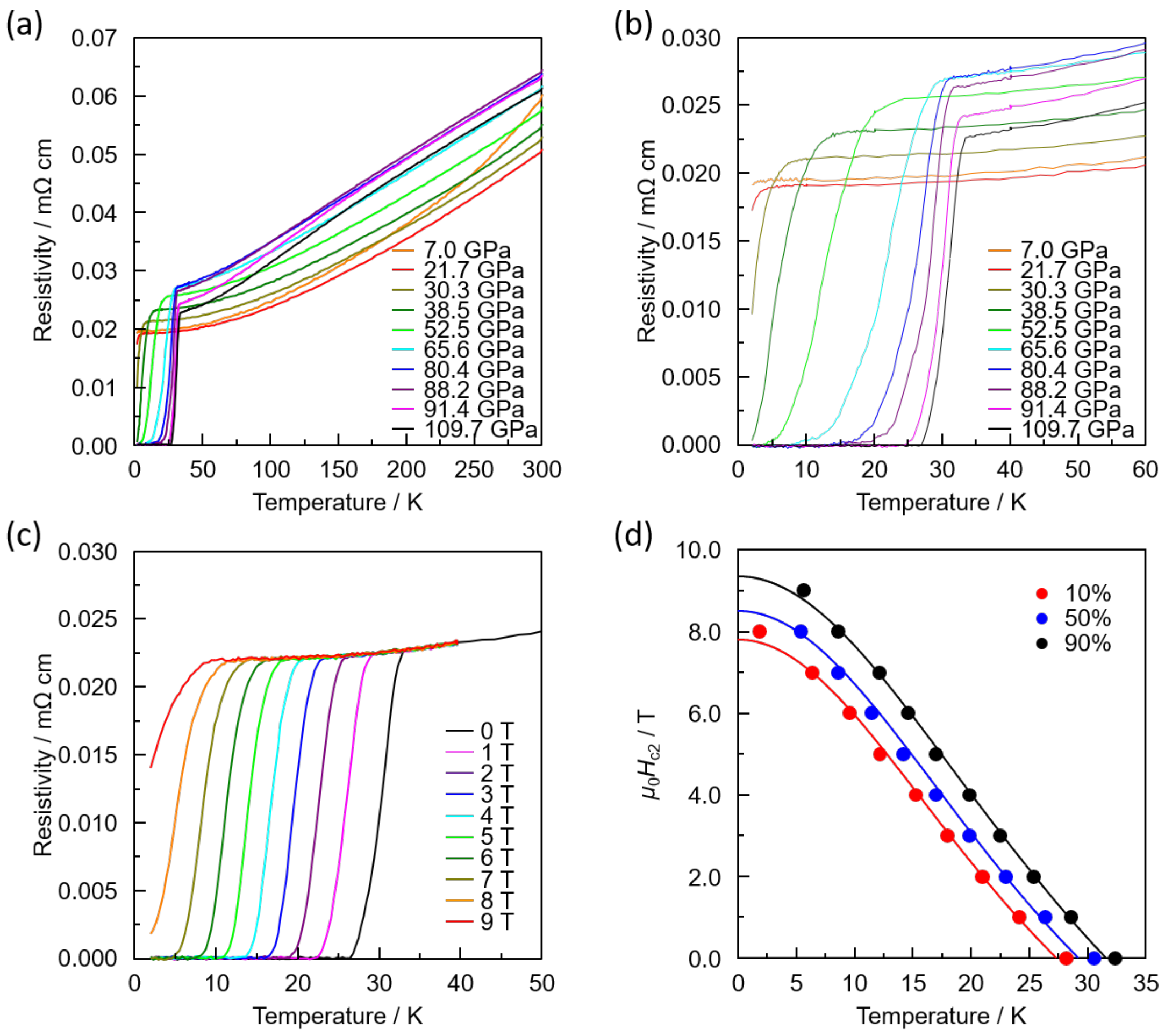

FIG. 2 (color online). Transport properties of $MoB_2$ as functions of pressure and magnetic fields in run IV. (a) Electrical resistivity $\rho(T)$ of $MoB_2$ as a function of temperature at different pressures. (b) Enlarged $\rho(T)$ curves in the vicinity of the superconducting transition. (c) $\rho(T)$ under various magnetic fields at 109.7 GPa. (d) Temperature dependence of upper critical field $\mu_0 H_{c2}(T)$ at 109.7 GPa. Here, the $T_c$s are determined at the 10%, 50%, and 90% of the normal state resistivity just above the onset superconducting transition temperature, respectively. The solid lines represent the fits using the Ginzburg-Landau formula.

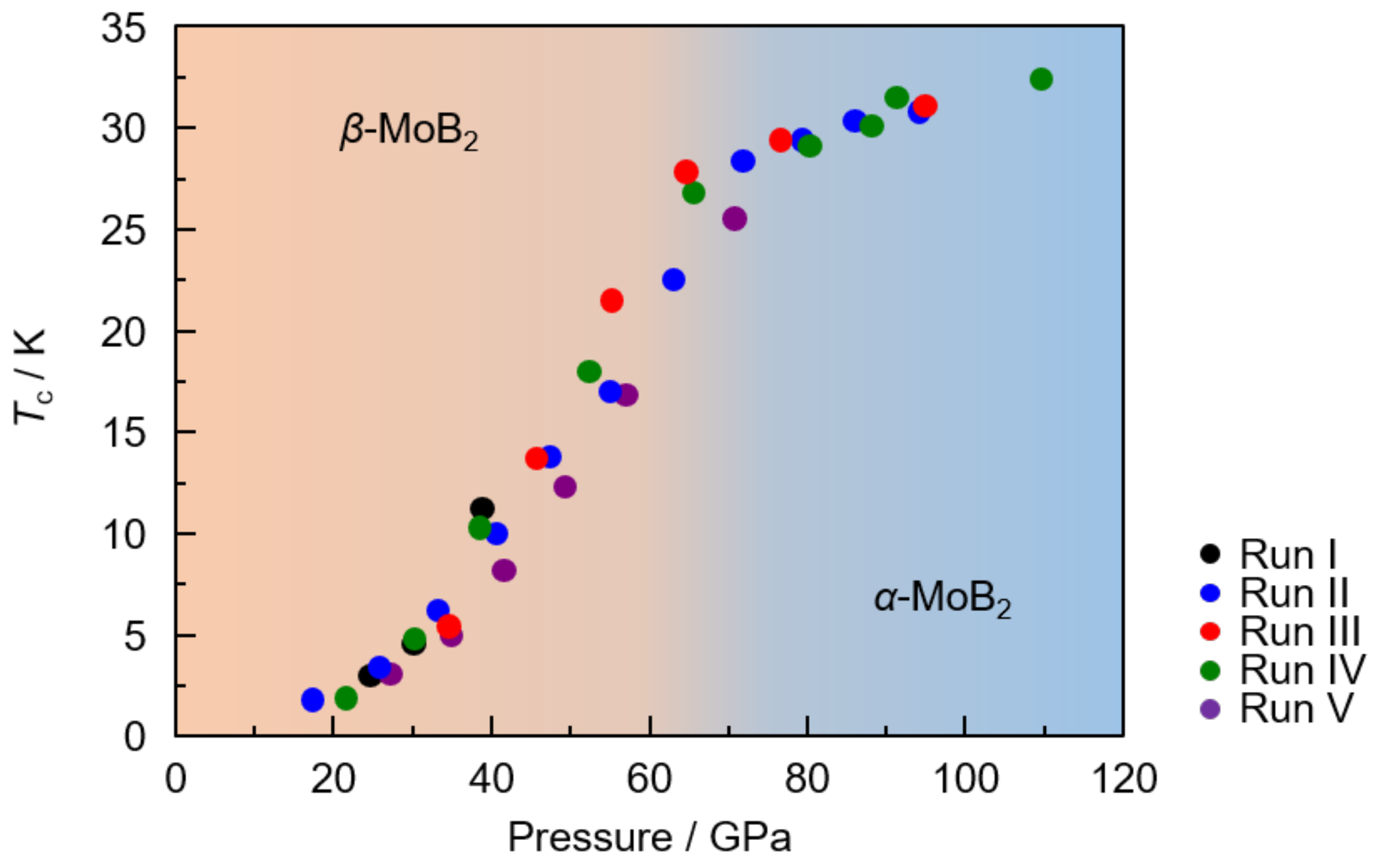

FIG. 3 (color online). Phase diagram of $MoB_2$. Pressure dependence of the superconducting transition temperatures $T_c$s for $MoB_2$ up to 109.7 GPa in different runs. The values of $T_c$ are determined from the high-pressure resistivity (90% $\rho_n$ criterion).

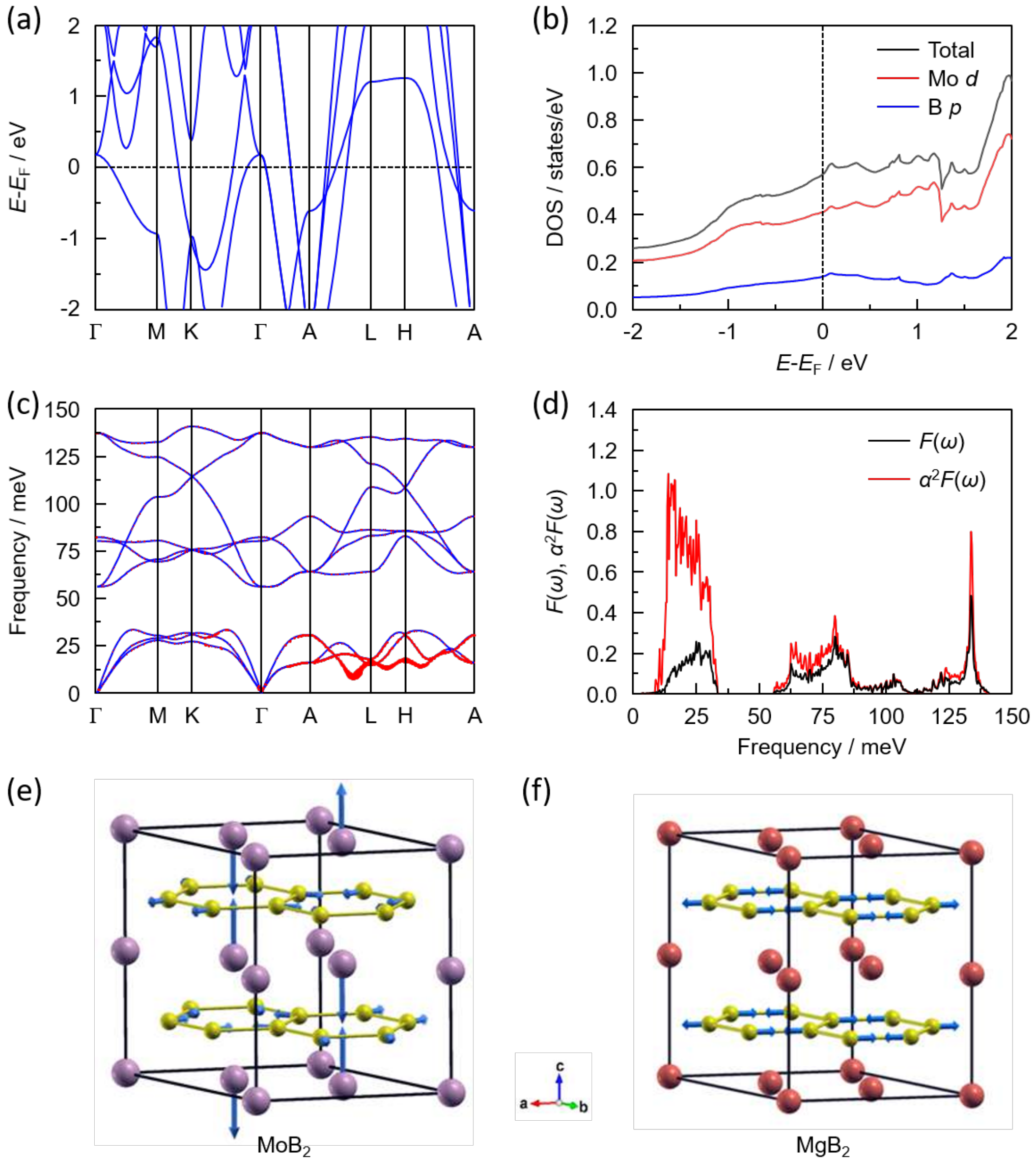

FIG. 4 (color online). Calculated electronic structure and atomic displacements for the typical phonon modes of $\alpha$-$MoB_2$ under 90 GPa. (a) Electronic band structure. (b) Density of states. (c) Phonon dispersion. The size of red dots schematically denotes the electron-phonon coupling strength $\lambda_{q\nu}$. (d) Phonon density of states $F(\omega)$ (black line) and Eliashberg spectral function $\alpha^2F(\omega)$ (red line). (e) Atomic displacements of the lowest acoustic mode at the H point (FIG. 4c), which has a large electron-phonon coupling (EPC). A $\sqrt{3} \times \sqrt{3} \times 2$ supercell is employed. The arrows on Mo/B atoms indicate vibrational directions and magnitudes, where the arrows on B atoms are magnified by two times for clear exhibition. (f) The in-plane B-B stretching mode for $MgB_2$ at ambient pressure. The Mo, Mg, and B atoms are represented by purple, red, and yellow balls, respectively.

Table 1. Typical of superconducting borides.

| Number | Compound | $T_c$ / K | Structure | Ref. |
|---|---|---|---|---|
| 1 | $MgB_2$ | 39 | $AlB_2$ | [1] |
| **2** | **$MoB_2$** | **32.4** | **$AlB_2$** | **This work** |
| 3 | $YPd_2B_2C$ | 23 | $LuNi_2B_2C$ | [33] |
| 4 | $LuRh_4B_4$ | 11.76 | $CeCo_4B_4$ | [34] |
| 5 | $LuRuB_2$ | 9.99 | $LuRuB_2$ | [35,36] |
| 6 | NbB | 8.25 | $\alpha$-TlI | [33] |
| 7 | $Mo_2BC$ | 7.5 | $Mo_2BC$ | [37] |
| 8 | $YB_6$ | 7.1 | $CaB_6$ | [38] |
| 9 | $Mo_2B$ | 5.07 | $\theta$-$CuAl_2$ | [39] |
| 10 | $ZrB_{12}$ | 5.82 | $UB_{12}$ | [38] |
| 11 | $Ba_{0.67}Pt_3B_2$ | 5.6 | $Ba_{0.67}Pt_3B_2$ | [40] |
| 12 | $Re_3B$ | 4.7 | $Re_3B$ | [41] |
| 13 | $LuOs_3B_2$ | 4.67 | $CeCo_3B_2$ | [42] |
| 14 | $YB_2C_2$ | 3.6 | $YB_2C_2$ | [43] |
| 15 | $Ru_7B_3$ | 2.58 | $Th_7Fe_3$ | [33] |